\documentclass[twocolumn]{aastex631}
\usepackage{subfigure}
\usepackage{amsmath}
\usepackage{hyperref}

\def\arcsec{{$^{\prime\prime}$}}

\def\Msun{\,{\rm M$_{\odot}$}}

\def\Lsun{\,{\rm L$_{\odot}$}}

\newcommand{\kms}{\mbox{km\,s$^{-1}$}}

\def\degr{$^{\circ}$}
\UseRawInputEncoding

\shorttitle{A Keplerian accretion disk around an early B-star }
\shortauthors{Sandell \& Vacca}

\begin{document}

\title{A rotating accretion disk around MWC\,297, a young  B1.5 Ve star}

\author{G\"oran Sandell}
\affil{ Institute for Astronomy, University of Hawai`i at Manoa,  Hilo,  640 N. Aohoku Place, Hilo, HI 96720, USA}
\email{gsandell@hawaii.edu}

\author{William Vacca}
\thanks{Visiting Astronomer at the Infrared Telescope Facility, 
which is \\ operated by the University of Hawaii under contract \\ 80HQTR19D0030
with the National Aeronautics and Space \\ Administration.}

\affil{NSFÕs NOIRLab, 950 N. Cherry Avenue, Tucson, AZ 85719, USA }
\email{bill.vacca@noirlab.edu}

\begin{abstract}
High resolution spectra with iSHELL on IRTF in the  K and M band of the young,
heavily accreting B 1.5e star MWC\,297 show numerous double-peaked CO lines.
These CO lines originate in an inclined gaseous disk in Keplerian rotation.
MWC\,297 is the only  early B star known to show a Keplerian disk in CO.
Analysis of the spectra show that $^{12}$CO 1 -- 0 is optically thick for the
low excitation lines. Even the $^{13}$CO 1 -- 0 and $^{12}$CO 2 -- 1 have
somewhat optically thick lines at low J levels. We find that the CO emission in
the disk can be  fitted with CO being in a narrow ring at a radius of 12 AU,
with a temperature of 1500 K, and a CO column density of 1.6 10$^{18}$
cm$^{-2}$. This model underestimates the line strength of high J lines,
indicating that they are excited by fluorescence. The CO overtone lines have a
similar temperature, The $^{13}$CO lines are much brighter than expected from
interstellar isotope ratios. The $^{13}$CO lines are wider than the $^{12}$CO
ones  suggesting different excitation conditions. The same is true for $^{12}$CO
 -- 1. We see strong absorption in $^{12}$CO and $^{13}$CO 1 -- 0 at low J
levels, which is due to two two cold foreground clouds. These clouds, one with a
temperature of 8.3 K and a column density of 6.7 10$^{17}$ cm$^{-2}$, and the
other one  colder and with lower column density,  can fully account for the
observed extinction toward MWC\,297.
\end{abstract}

\keywords{Star formation (1569); Star forming regions (1565); Massive stars (732); 
Circumstellar disks (235); Infrared astronomy (786)}

\section{Introduction}

Most young, low mass stars are surrounded by circumstellar disks. The same
is also true for intermediate mass stars up to a spectral type of $\sim$B8
\citep{Williams11}. Direct detections of disks around  early B  or O stars  are
rare. This does not mean that high mass stars do not have accretion disks;
rather it indicates that disks are much more short-lived around high mass stars
so that by the time these stars become optically visible they have already
dispersed their accretion disks.  Furthermore, high mass stars are generally
more distant than low mass stars, making it harder to spatially resolve their
accretion disks. Even O stars with stellar masses of 20 -- 30 \Msun\ appear to
have disks in their heavy accretion phase \citep[see
e.g.,][]{Ilee13,Beltran16,Zapata19,Sandell20,Moscadelli21}.  However, such
stars are difficult to study because they are always in clusters and deeply
embedded.

Most of the evidence for disks around high mass stars comes from deeply embedded
young objects observed with interferometric means, mostly in the
millimeter/submillimeter regime with ALMA and NOEMA and other array telescopes.
There are also some detections at optical/infrared wavelengths
\citep{Beltran16}. There has been some success in looking for Keplerian rotation
in accretion disks using the rovibrational CO lines at 2.3 $\mu$m and at 4.5 -
.2 $\mu$m. CO overtone emission ($\upsilon$ = 2 -- 0 and $\upsilon$ = 3 -- 1)
was first detected in the BN object (mass $\sim$ 10\Msun{}) \citep{Scoville79},
and in the early B-star MWC\,349\,A \citep{Kraus00}. The latter has a mass of
$\sim$ 30 \Msun, although a recent paper by \citet{Kraus20} argue that it
is a B[e] supergiant. There have been a number of detections of bandhead
emission toward massive young stellar objects
\citep{Ishii01,Blum04,Bik04,Bik06,Wheelwright10,Cooper13,Ilee13, Pomohaci17}.
For these objects the stellar mass is usually  deduced from the observed
bolometric luminosity. These can be highly uncertain because it includes
emission from the accretion process and possible emission from nearby objects,
resulting in an overestimate of stellar mass, see, e.g., W\,33\,A, for which
\citet{Pomohaci17} determine a mass 17.2 \Msun, while \citet{Navarete21} find a
mass of 10\Msun. Yet it is clear that some of these are indeed very young high
mass stars. The CO fundamental ($\upsilon$ = 1 - 0) rovibrational lines,
located in the M band, 4.5 -- 5.2 $\mu$m, have  been detected in low mass stars
\citep{Najita03}, HAEBE stars, and transitional disks objects
\citep{Brittain03,Blake04,Salyk09,Plas15,Doppmann17,Banzatti22}, but not  in
high-mass stars (M $>$ 8 \Msun{}), except for MWC\,297, the subject of this
study.

MWC\,297 is a bright, nearby HAeBe star with a spectral type B1.5 Ve
\citep{Drew97} and a mass of $\sim$ 10 \Msun, located at a distance of 418 pc
\citep[Gaia DR3,][]{Riello21}. It has been the subject of several  studies with
ESO's Very Large Telescope Interferometer (VLTI)
\citep{Acke08,Weigelt11,Lazareff17,Kluska20}. Most of these studies suggest that
the star is surrounded by an almost face-on disk. The star, however, drives an
ionized outflow, with spatially separated outflow lobes, and therefore cannot be
face-on \citep{Sandell11}. Mid-infrared imaging with FORCAST on SOFIA 
\citep{Vacca22} shows that hot dust traces the outflow lobes at wavelengths $>$
 $\mu$m. Simple geometrical  modeling the mid-infrared morphology constrain
the disk inclination rather well and give an inclination angle i = 55\degr.

Here we present and discuss high resolution M band data obtained with the iSHELL
instrument at the IRTF. The spectrum reveals double-peaked CO emission lines,
which we have modeled with a rotating Keplerian disk to determine the properties
of the emission region. We also discuss and analyze the absorption spectra from
the cold foreground cloud, which are seen in low $\upsilon$ = 1 -- 0 P and R
transitions. We note that MWC\,297 was included in a large survey of planet
forming disks observed with the same instrument and the same wavelength range,
and these data have recently been published \citep{Banzatti22}. While 
\citet{Banzatti22} did classify the double peaked CO profiles in MWC\,297 as
originating in a Keplerian accretion disk, they provide very few details. Here
we present a more comprehensive analysis of of both CO overtone emission at 2.3
$\mu$m as well as the rovibrational spectra in the M-band.

\section{iSHELL observations and data reduction}

Observations of MWC\,297 were obtained at the NASA Infrared Telescope Facility
(IRTF) on Mauna Kea on 2020 Sep 30 (UT) and on 2022 April 26 (UT) with
iSHELL, the facility near-infrared high resolution cross-dispersed spectrograph
\citep{Rayner22}. The M band observations used the M1 setting of iShell 
both in 2020 and 2022. This mode yields spectra spanning the wavelength range
.5 -- 5.2 $\mu$m over 16 spectral orders. The observations were acquired in
``pair mode", in which the object was observed at two separate positions along
the 15\arcsec-long slit. The slit width was set to 0\farcs375, which yields a
nominal resolving power of 88,000 for the spectra. (At the distance of MWC\, 297
of 418 pc, the iShell 0\farcs375 slit spans $\sim 160$ AU.) Twenty individual
exposure of MWC\, 297, each lasting 28 s, were obtained using the M1 setting of
iShell. In 2020 the slit was set to the parallactic angle (46.0\degr{}) during
the observations. In 2022 the observations were done using two slit
positions, 197\degr\ and 17\degr, which were coadded in the post reduction. 
The observations both times were acquired in ``pair mode", in which the object
was observed at two separate positions along the 15\arcsec-long slit. The slit
width was set to 0\farcs375, which yields a nominal resolving power of 88,000 for
the spectra.  At the distance of MWC\, 297, 418 pc, the iShell 0\farcs375 slit
spans $\sim 160$ AU. Long observations of an A0$\, $V star, used as a ``telluric
standard" to correct for absorption due to the Earth's atmosphere and to flux
calibrate the target spectra, were obtained immediately prior to the
observations of MWC\, 297.  In 2022 we also observed MWC\,297 in the K
band. These observations were done in the K3 setting, providing a spectral
coverage from 2.256 to 2.55 $\mu$m.  A set of internal flat fields and arc
frames were obtained immediately after the observations of MWC\, 297 for flat
fielding and wavelength calibration purposes.

The data were reduced using Spextool \citep{Cushing04}, the IDL-based package
developed for the reduction of SpeX and iShell data. The Spextool package
performs non-linearity corrections, flat fielding, image pair subtraction,
aperture definition, optimal extraction, and wavelength calibration. The sets of
spectra resulting from the individual exposures were median combined and then
corrected for telluric absorption and flux calibrated using the extracted A0\, V
telluric standard spectra and the technique and software described by
\citet{Vacca03}. The spectra from the individual orders were then spliced
together by matching the flux levels in the overlapping wavelength regions, and
regions of poor atmospheric transmission were removed. However, for the
September 2020  the final M spectrum of the A0V star had relatively poor S/N (on
the order of 30), and this limited the final S/N of the spectrum of MWC\,297,
which is considerably brighter in the M band. (The reduced spectrum of MWC\,297
has a S/N of several hundred.) Therefore, an alternate telluric correction was
derived using the {\it xtellcor\_model} software developed by \citet{Boogert22}
and then applied to the reduced spectrum of MWC\,297. Although this was
effective at correcting the telluric absorption, the resulting spectrum was
still uncalibrated. Therefore, we fit a polynomial to the continuum of both the
A0V-corrected spectrum and the model-corrected spectrum, and multiplied the
model-corrected spectrum by the wavelength-dependent ratio of the polynomials.
This yielded a telluric-corrected and flux-calibrated spectrum of MWC\,297. The
S/N varies across the spectrum but is on the order of several hundred across the
entire M1 wavelength range. For the 2022 data we used a brighter A0$\, $V
star, which could directly be used for flux calibration. Direct comparison of
the two calibrated data sets show that the calibration agrees within 10\% for
both data sets and we have therefore added together the two data sets, after
correcting both spectra for Doppler shifts due to the relative motion of the
earth and the source, and extracted flux densities for all lines not affected by
telluric absorption. In 2020 the Doppler shift for MWC\,297 was -11.5 \kms,
which severely affected CO lines with J values $\lesssim$ 15, while it was +41.2
\kms\ in 2022, which largely shifted all CO lines away from telluric absorption
features.

\section{A short overview of what has been learned from studies of rovibrational CO}

Although CO vibrational overtone emission was first detected in the inner disks
of both high and low mass objects, but it is not as ubiquitous as the CO
fundamental  ($\Delta\upsilon$ = 1) transitions of CO at 4.6 $\mu$m. The
detection rate of overtone emission in  unbiassed surveys is about 25\%. In part
this is due to the relatively low A-values, which require high column densities
of hot gas for the lines to be detectable, but other factors play a role as
well. Modeling by \citet{Ilee18} suggest that moderate accretion rates produce
the most prominent and hence detectable CO first overtone emission. If the
accretion rate is too high ($\gtrsim$ 10$^{-4}$ \Msun{}~yr$^{-1}$), it results
in large dust sublimation radii, a larger contribution of hot dust to the K band
continuum and therefore a lower CO to continuum ratio. Low accretion rates
($\lesssim$ 10$^{-6}$  \Msun{}~yr$^{-1}$) however, result in smaller dust
sublimation radii, a smaller  CO emitting area and therefore a lower CO to
continuum ratio. Although CO overtone emission is detectable in MWC\,297, which
has an accretion rate of  3 10$^{-4}$ \Msun{}~yr$^{-1}$, the first overtone band
is barely visible in high resolution spectra, suggesting that the gas is relatively cold, or
about 1000 K.  The higher vibrational levels are filled with
double peaked spectra. This will be discussed
later in the paper.

The rovibrational CO line emission at 4.5 -- 5.2 $\mu$m is as already mentioned
much more ubiquitous and seen in a variety of objects, ranging from Transition
Objects to Classical T Tauri and HAEBE stars, but contrary to CO overtone
emission it is not seen in high-mass stars except for MWC\,297 \citep[and this
paper]{Banzatti22} and perhaps in BN \citep{Scoville83}. \citet{Banzatti22} gives
an up to date list of all the high resolution K and M-band  CO surveys done over the
last 20 years. Therefore there is no reason for us to repeat it here. Instead we
try to surmmarize the major findings resulting form these studies and
complementary theoretical modeling.

The rovibrational CO emission at 4.7 $\mu$m  is very common in protoplanetary
disks and arises in the inner 0.1 - 2 AU region in disks around low mass stars
\citet{Najita07} and up to 30 AU or more in HAEBE stars
\citep{Plas15,Banzatti18, Banzatti22}. The line shapes vary
\citep{Bast11,Brown13,Banzatti15, Banzatti22}. They can be narrow and single
peaked, narrow with a broad pedestal, broad but single peaked, or pure double
peaked Keplerian profiles. Some stars also show the CO in absorption, which in
most cases originates from colder foreground gas. Double peaked profiles, which
are expected to be a true disk signature, are surprisingly uncommon
\citep{Bast11}, although in the large survey by \citet{Banzatti22} the detection rate of 
double peaked lines was quite common, $\sim$ 50\%.

\citet{Banzatti22} define the line profiles in two broad groups
depending on line shape: triangular and double peaked lines. In their
classification the  single peaked narrow lines with a broad pedestal and the
broad, but single peaked lines all fall into this category. These lines are
often somewhat asymmetric with the blue-shifted side being stronger
\citet{Herczeg11,Brown13}. \citep{Banzatti22} define a line shape parameter S as
 S = FW10\%/FW75\%, where FW stands for Full Width at 10 and 75\%, respectively.
Lines with a shape value  $\gtrsim$ 2.5 are triangular, and $\lesssim$ 2.5 they
are generally double peaked Keplerian or narrow single peaked lines. The latter
are believed to represent Keplerian disk emission seen in face on disks. The
single peaked narrow lines with a broad pedestal, i.e. triangular shaped lines
cannot be explained by Keplerian motion. The broad pedestal may still be
Keplerian, but the narrow component appears to originate in a slow mowing disk
wind \citet{Bast11,Brown13}. \citet{Pontoppidan11} found  from spectroastrometry
of narrow line with a broad pedestal that the narrow line emission is too
compact and asymmetric to originate in a Keplerian disk. The broad single peaked
lines are also believed to be a combination of a Keplerian disk rotation and a
disk wind, but the disks are seen with higher inclination angles, producing
broader lines \citep{Brown13}. 

Line shape alone does not really discriminate where the CO lines are formed,
whether they originate from the disk or the outflow or both. Spectroastrometry
\citep{Pontoppidan11, Brittain18} can achieve subarcsecond resolution and hence
put spatial constraints on the emitting region, which has been used to
discriminate between Keplerian and wind models. Another powerful constraint is
the strength of higher vibrational levels, which have been detected up to
$\upsilon$ = 6 -- 5 in HAEBE stars \citep{Brittain07,Jensen21,Banzatti22}. For
double-peaked lines, the work by \citet{Banzatti18,Banzatti22,Bosman19} suggests
that one can identify three  emitting regions that show a dependence on the
$\upsilon$2/$\upsilon$1 ratio depending on the location of their emitting radii,
$R_{\rm CO}$, compared to their dust sublimation radius, $R_{\rm subl}$. Regions
a and 1b both have high vibrational  ratio $\upsilon$2/$\upsilon$1 =  0.2 --
.0, but come from very different regions in the disk, although both are largely
dust free. Region 1a is inside $R_{\rm subl}$, while 1b is inside a dust cavity
observed by millimeter interferometry. Region 1b therefore has lower rotational
temperatures and narrower line widths and it appears that UV fluorescence
largely excites the higher vibrational levels in this region \citep{Bosman19}.
In region 2 the vibrational ratio $\upsilon$2/$\upsilon$1 =  0.01 -- 0.2, and
the CO is emitted from regions outside the dust rim.

\section{Analysis and Results}
\label{sect-Analysis}

\begin{figure}[t]
\includegraphics[width=7cm,angle=90]{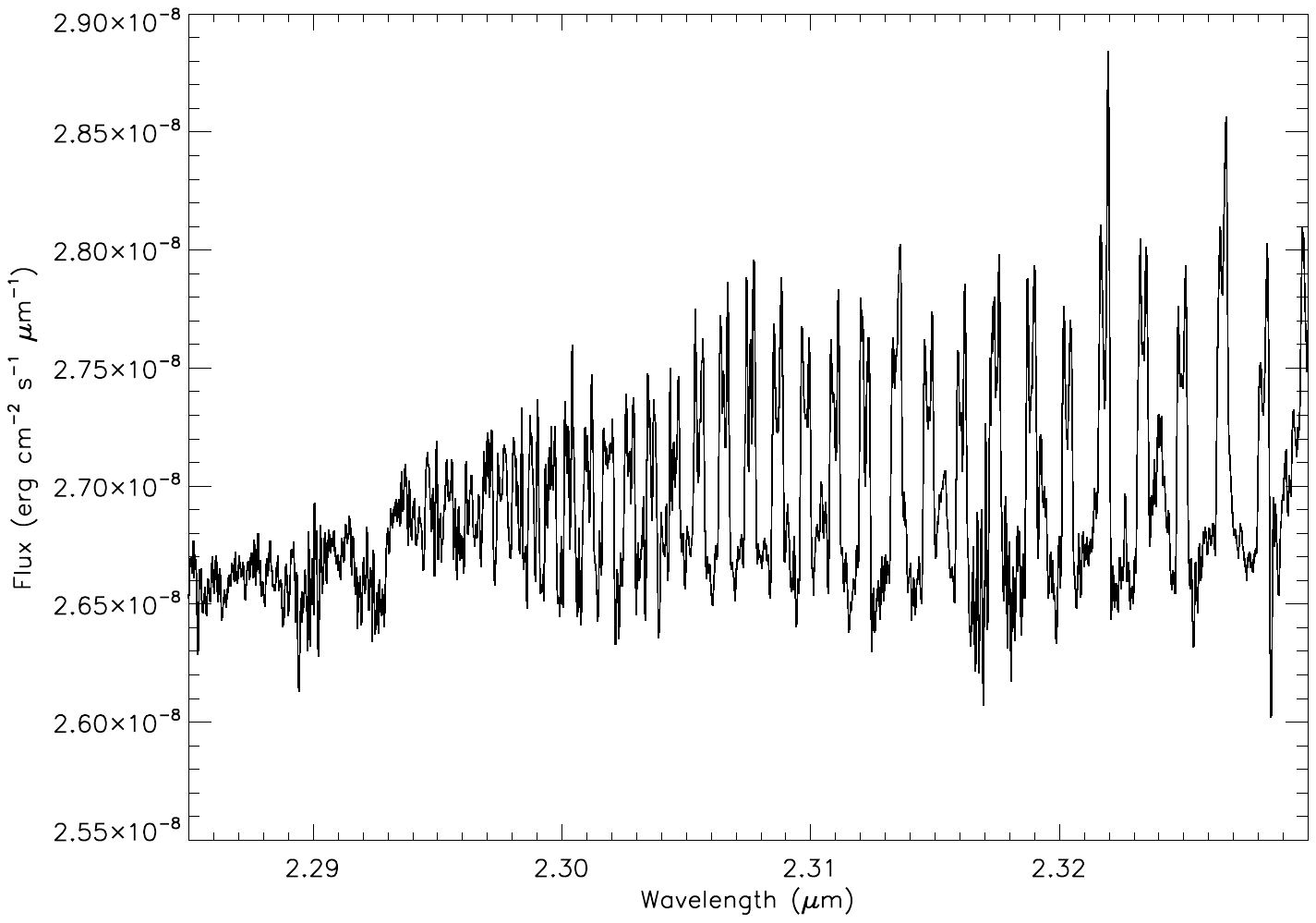}
\figcaption[]{The CO overtone bands observed with iSHELL in K band showing a large number 
of double peaks lines. The 2 -- 0 bandhead at 2.29 $\mu$m  is very weak.
\label{fig-bandheads}} 
\end{figure}

The iSHELL spectrum exhibits a large number of double-peaked emission lines from
rovibrational transitions of CO. The $\upsilon$ = 1 -- 0  CO lines up to
rotational level P(47) and higher vibrational levels up to $\upsilon$ = 5 -- 4
are clearly seen. Emission from $^{13}$CO is also seen, in both the $\upsilon$ =
 -- 0, and $\upsilon$ = 2 - 1 transitions. The profiles of the lowest
rotational transitions exhibit strong narrow absorption lines, due to absorption
from a cold foreground cloud, which cuts out a substantial fraction of the
emission. Three H lines (Pf $\beta$, Hu $\epsilon$, and Hu $\delta$) are also
seen in emission. The K band iSHELL spectrum also shows a large number of double 
peaked lines, all from CO overtone bandhead emission, see Figure~\ref{fig-bandheads}. 

\begin{figure}[t]
\includegraphics[width=8.5cm,angle=0]{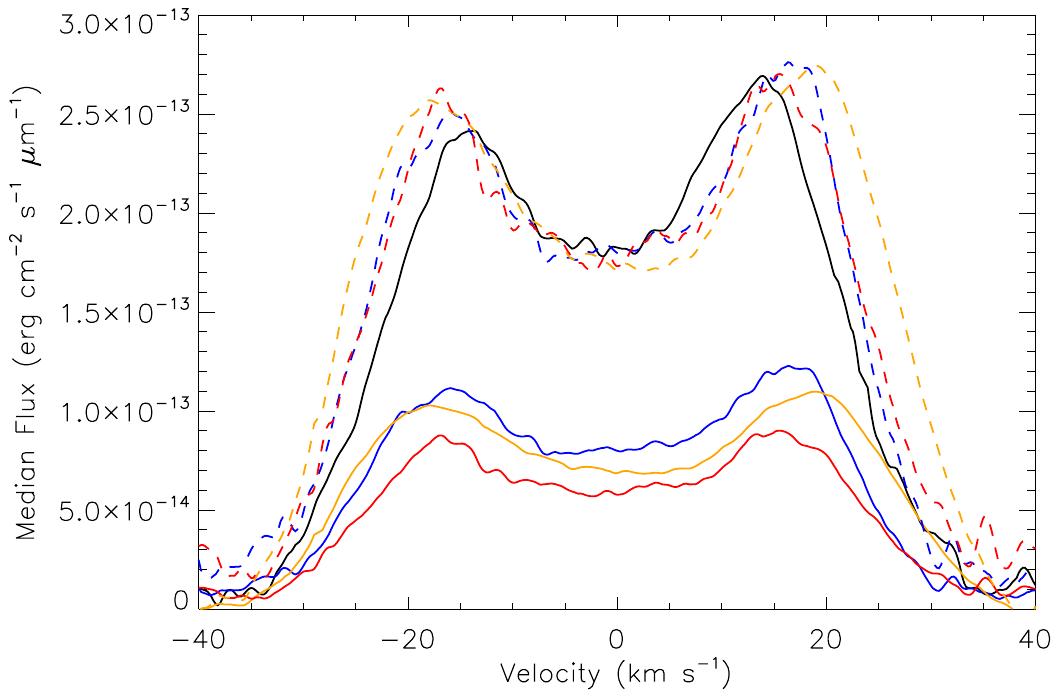}
\figcaption[]{Median line profiles of  $^{12}$CO $\upsilon$ = 1 -- 0 and 2 -- 1 and $^{13}$CO
 -- 0  {\bf and  $^{12}$CO 2 -- 0} emission lines.
The dashed blue line is the 
$^{12}$CO 2 -- 1 profile scaled by 2.25, the dashed red line is the $^{13}$CO 1 -- 0 profile scaled
by 3, the dashed orange is the scaled  $^{12}$CO 2 -- 0. The scaled profiles demonstrate the differences 
in widths compared to the $^{12}$CO
--0 profile. \label{fig-COprofiles}} 
\end{figure}

It is clear that the double peaked $^{12}$CO profiles must originate in a
rotating accretion disk. To enable a better comparison between $^{12}$CO and
$^{13}$CO lines we generated median line profiles of the $^{12}$CO $\upsilon$ =
 -- 0, $^{13}$CO 1 -- 0  and  also for the $^{12}$CO 2 -- 1  and $^{12}$CO 2 -- 0 emission
lines from the spectrum after subtracting a low-order polynomial continuum.
Since we see no difference between low and high J lines we created median lines
profiles of all unaffected by absorption or blending. For the $^{12}$CO 1 -- 0
we used R(14), R(9), R(6), P(8), P(9), P(12), P(14), P(26), P(34), (P37), P(38),
P(45), and P(47). For   $^{13}$CO 1 -- 0 we used R(29), R(24), R(22), R(21),
R(19), R(16), R(15), R(13), R(12), R(10), R(9), R(3), P(3), P(7), P(9), P(11),
P(12), P(16), P(17), and P(21). {\bf The} $^{12}$CO 2 -- 1 transitions were
R(27),  R(21), R(20), R(17), R(14), R(13), R(12), R(9), R(1), P(1), P(2), P(6),
P(8), P(15), P(21), P(25), P(26), P(28),  P(32), and P(33), while the transitions
for $^{12}$CO 2 -- 0 were R(6), R(7), R(11), R(12), R(14), R(17), R(18), R(20), R(21), R(22), 
R(23), and  R(24).
These were
re-interpolated and centered at the theoretical line center, see
Figure~\ref{fig-COprofiles}.  From this plot it is easy to that the $^{13}$CO 1
-- 0   lines are broader and the flank of the $^{13}$CO lines also appear
somewhat steeper.  The $^{12}$CO 2 -- 1 lines are also broader than the 1 -- 0
lines. The overtone $^{12}$CO 2 -- 0 lines are even broader than any of the rovibrational lines. 
The separation between the line peaks are also wider for both $^{13}$CO 1
-- 0 and $^{12}$CO 2 -- 1 compared to $^{12}$CO 1 -- 0 lines indicating that
these lines are inward of the disk relative to the location of the $^{12}$CO 1
-- 0 emitting gas.  In Table~\ref{tab_parameters} we give the measured 
 FWHMs and peak separations for all the lines shown in Fig.~\ref{fig-COprofiles}.

At 10\% level the line widths are similar or $\sim$ 65 \kms, suggesting that the
 inner radius is similar for all of them. The FW10\% maybe slightly larger
for the $^{12}$CO 2 -- 0 lines. We measure a linewidth of 68 \kms. It is not
clear whether this difference is real or simply due to measurement inaccuracy,
especially since \citet{Banzatti22} quote a FW10\%  of 68.9 \kms. They also
quote a slightly larger FWHM of  46.8 \kms\ for $^{12}$CO 1 -- 0 than what we
find, 45.3 \kms .  We determine a V$_{lsr}$ = +6.7 $\pm$ 2.2 \kms for the
emission lines. This is a weighted mean of the September 2020 and April 2022
data sets.  \citet{Banzatti22} find a center velocity of + 7 \kms\ (converted to
LSR). which agrees very well with what we find.

\begin{table}[h]
\begin{center}
\caption{Line profile parameters for median averaged emission lines. 
The FWHM values have measurement  errors of $\sim$ 0.5 \kms, the errors for the peak 
separations are $\sim$ 1  \kms.\label{tab_parameters}} 
\begin{tabular}{lcc}
\hline\hline

Profile& FWHM &  Peak separation \\
 & [\kms{}] & [\kms{}]\\
 $^{12}$CO $\upsilon$  = 1--0 &  45.3  & 29 \\
 $^{12}$CO $\upsilon$  = 2--1 &  48.2 &  32 \\
 $^{13}$CO $\upsilon$  = 1--0 &  48.2    & 32\\
$^{12}$CO $\upsilon$  = 2--0 &  54.7   &  37 \\
\hline
\hline
\end{tabular}
\end{center}
\end{table}

\subsection{Modeling the emission lines with a rotating Keplerian disk.}
\label{sect-diskmodeling}

\begin{figure}[t]
\includegraphics[width=7.2cm,angle=90]{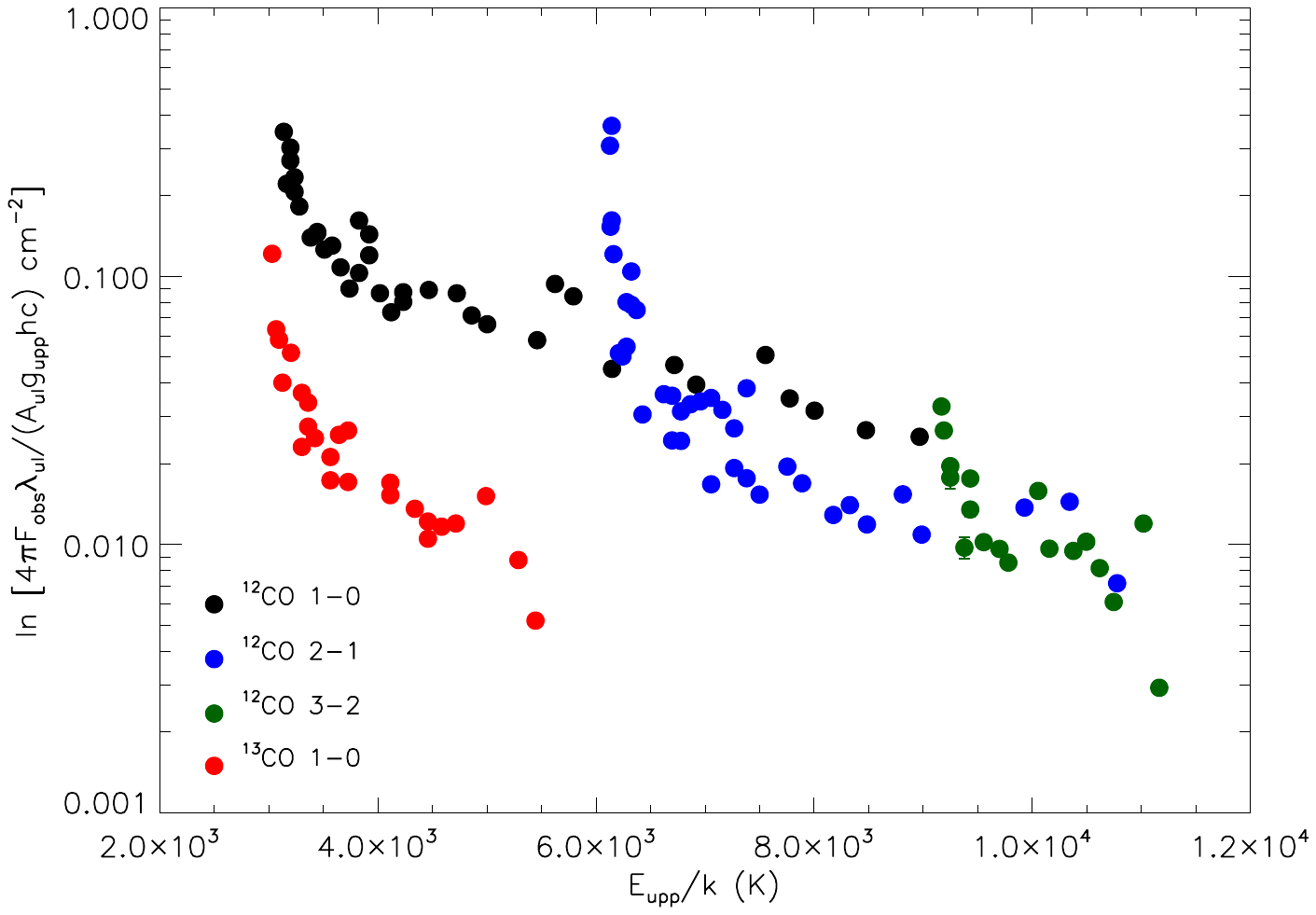}
\figcaption[]{Rotational diagram with the observed data points plotted with large dots. The 
line bands are identified by their colors in the bottom left part of the plot.  For
almost all of the data, the error bars are smaller than the dot size.
\label{fig-rotdiagram}
}
\end{figure}

\begin{figure}[t]
\includegraphics[width=7cm,angle=90]{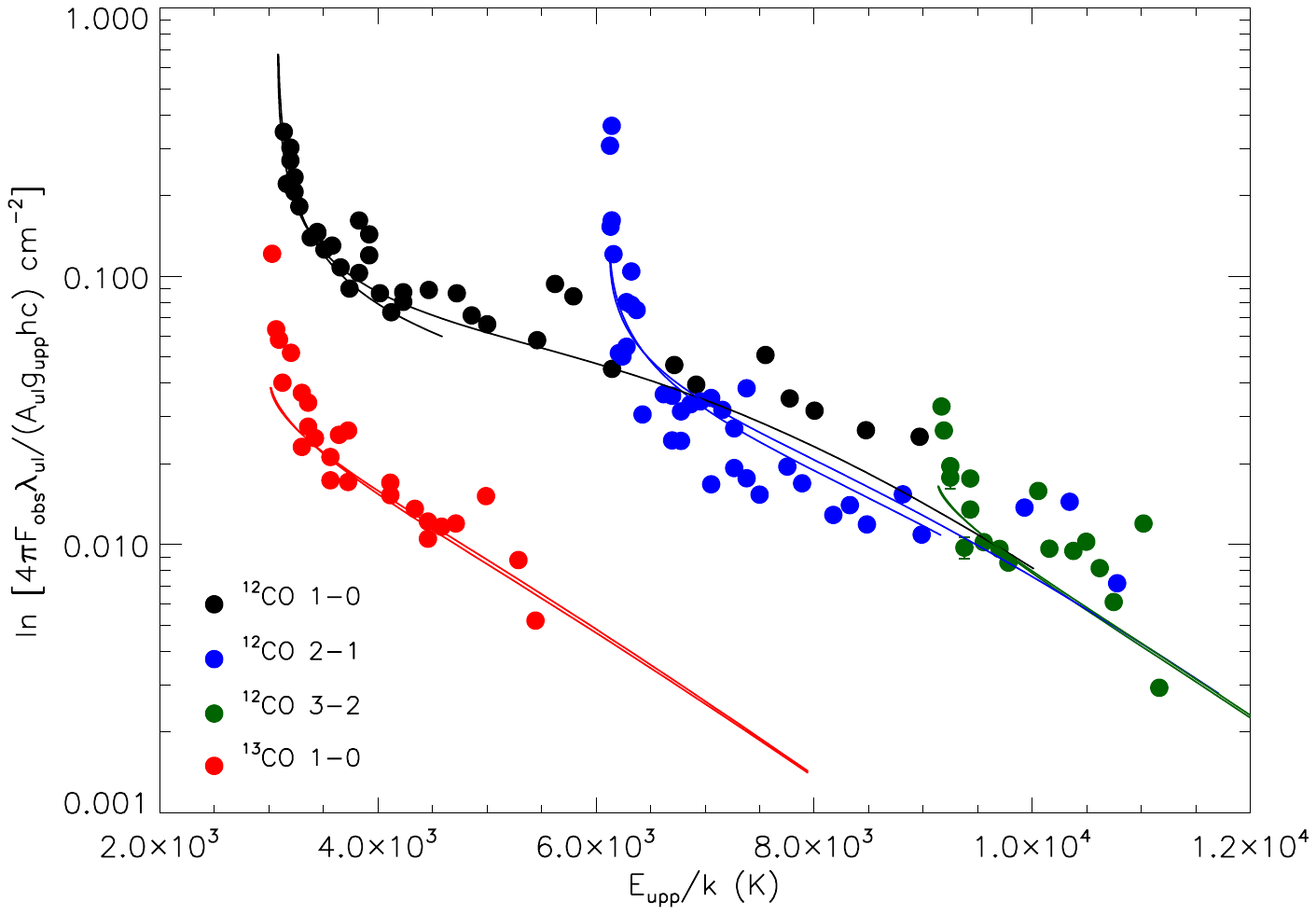}
\figcaption[]{ The ring model for a rotational temperature of 1500 K and a density of $1.6\times 10^{18}$
cm$^{-2}$ plotted with solid lines  on the data points plotted in
Figure~\ref{fig-rotdiagram}. The model matches the 1 -- 0 data quite well except
at the highest energies where it  underestimates the line strength. It also
underestimate the line strength at the higher vibrational states,  suggesting
that the excitation of these lines are dominated by fluorescence. The model overestimates the
 -- 1 data, most likely because its emitting area is smaller than that of 1 -- 0. The model calculations show
that the  $^{13}$CO abundance needs to be increased by a factor of 6 - 7 compared to normal interstellar abundances.
\label{fig-ladderfit} 
}
\end{figure}

\begin{figure*}
 \centering
  \begin{minipage}[]{8.3cm}{}
\includegraphics[width=6.9cm,angle=90]{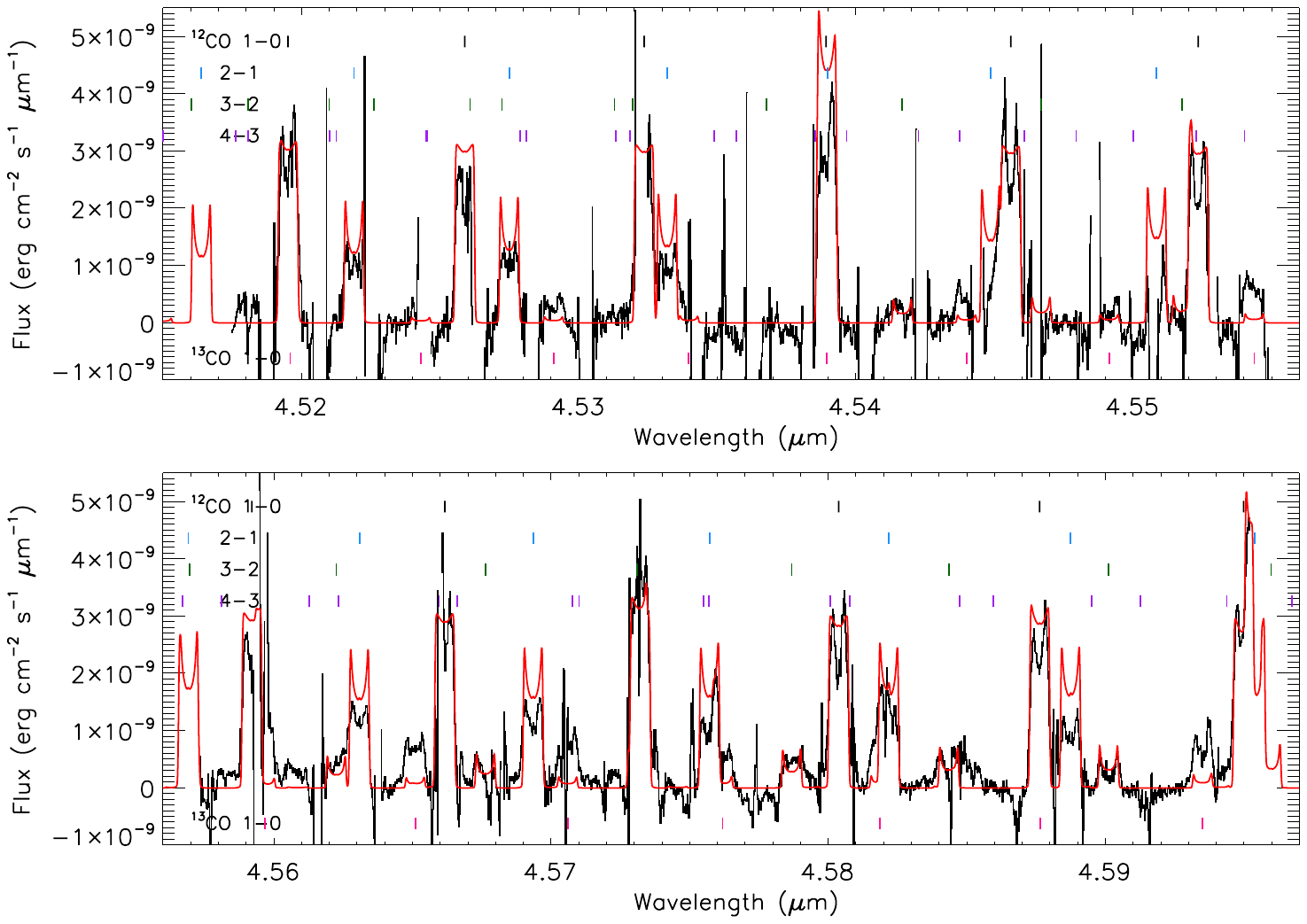}  
   \end{minipage}
    \begin{minipage}[]{8.3cm}{}
   \includegraphics[width=6.9cm,angle=90]{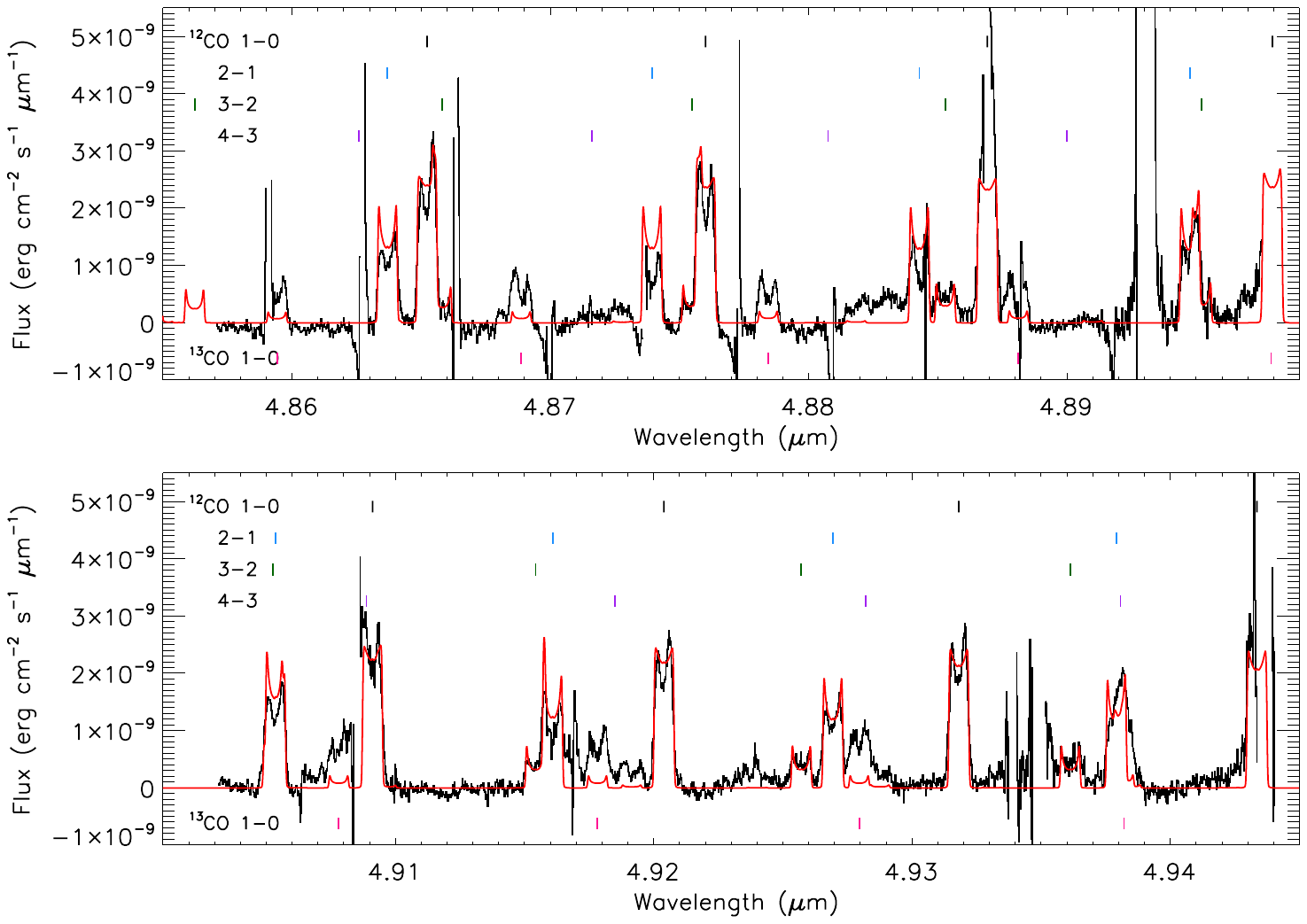}  
 \end{minipage}\\[-1.2cm]
 \begin{minipage}[]{8.3cm}{}
\includegraphics[width=6.9cm,angle=90]{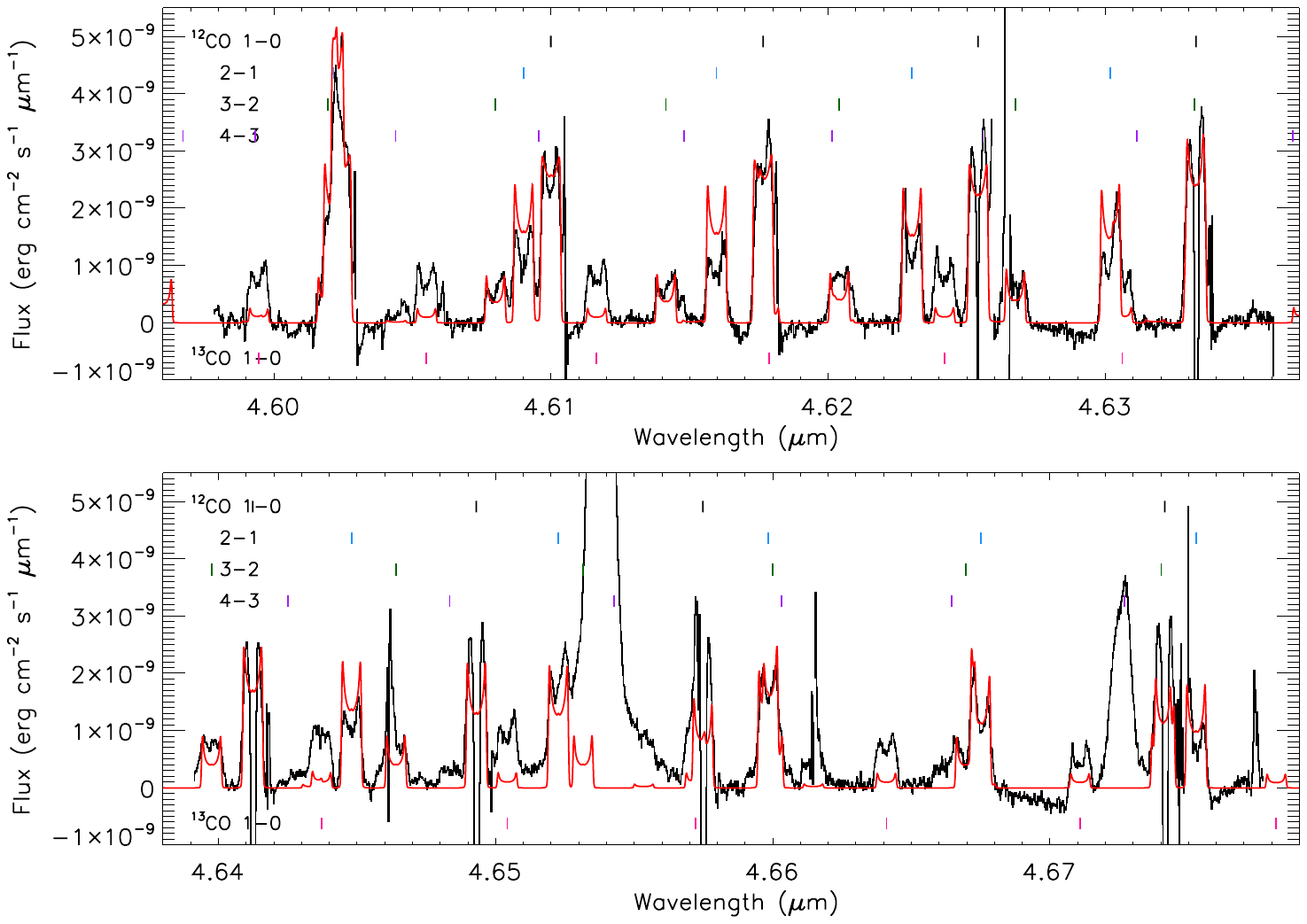}  
 \end{minipage}
  \begin{minipage}[]{8.3cm}{}
\includegraphics[width=6.9cm,angle=90]{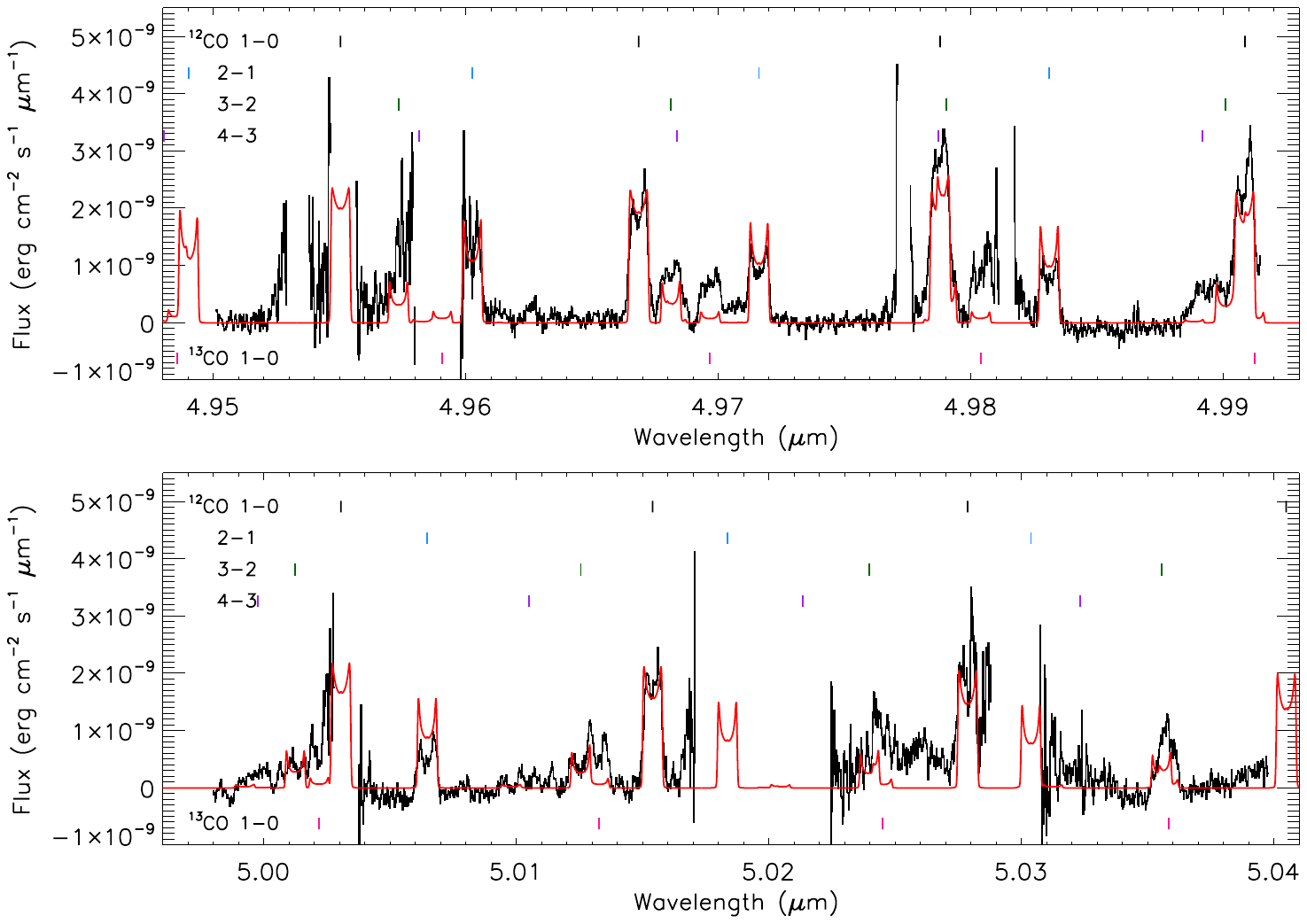}  
 \end{minipage}\\[-1.2cm]
\begin{minipage}[]{8.3cm}{}
\includegraphics[width=6.9cm,angle=90]{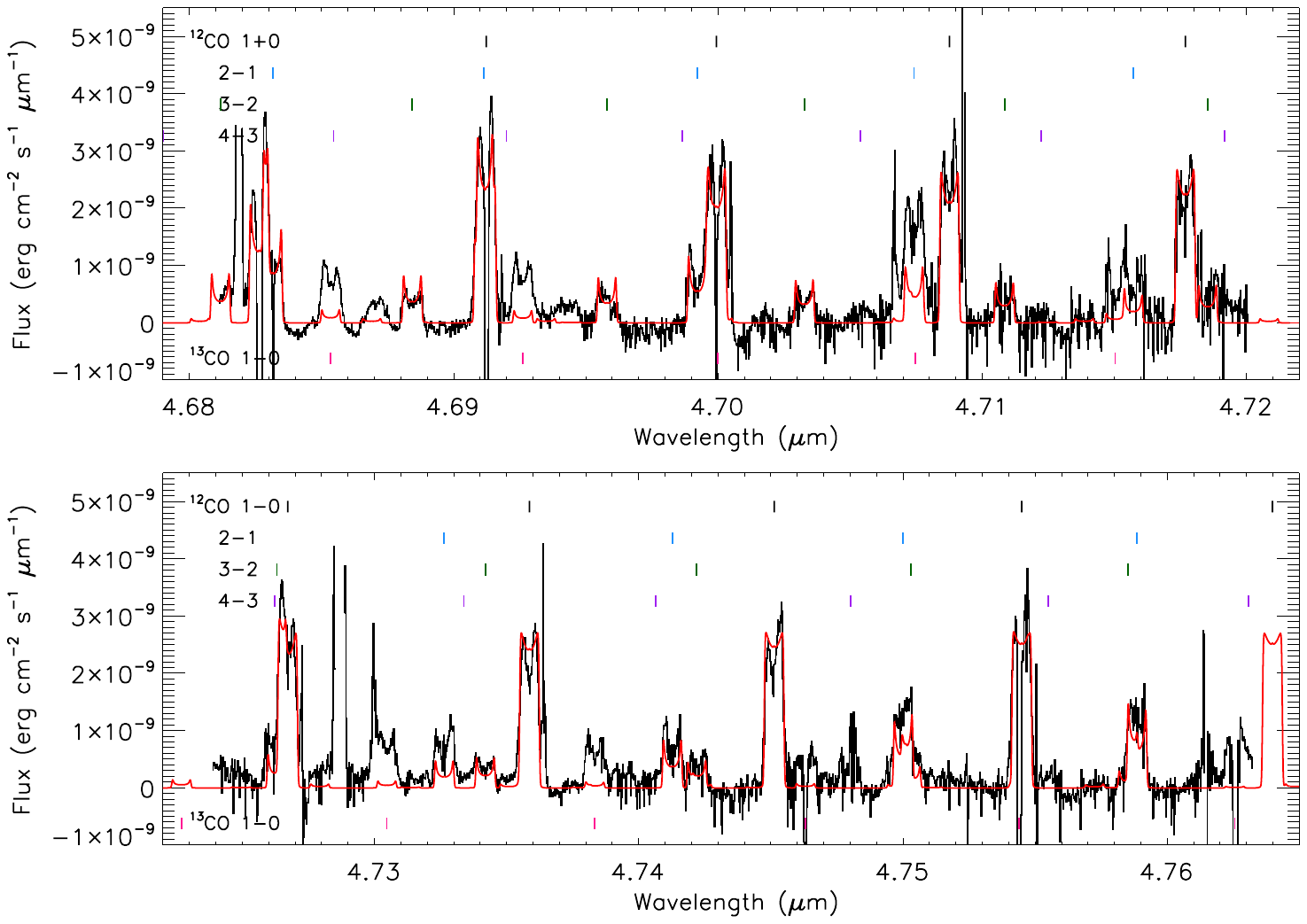}  
\end{minipage}
\begin{minipage}[]{8.3cm}{}
\includegraphics[width=6.9cm,angle=90]{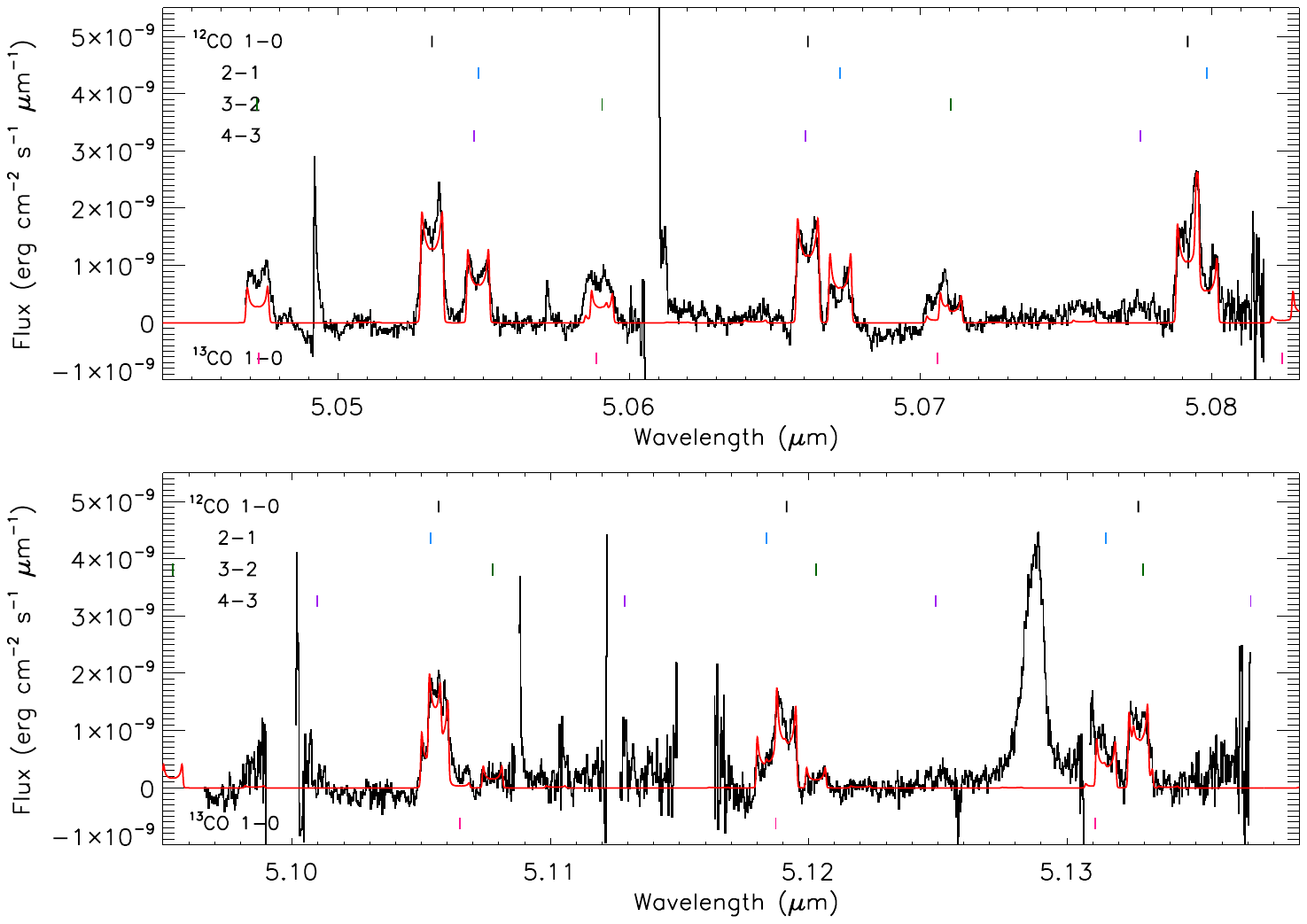}  
\end{minipage}\\[-1.2cm]
 \begin{minipage}[]{8.3cm}{}
\includegraphics[width=6.9cm,angle=90]{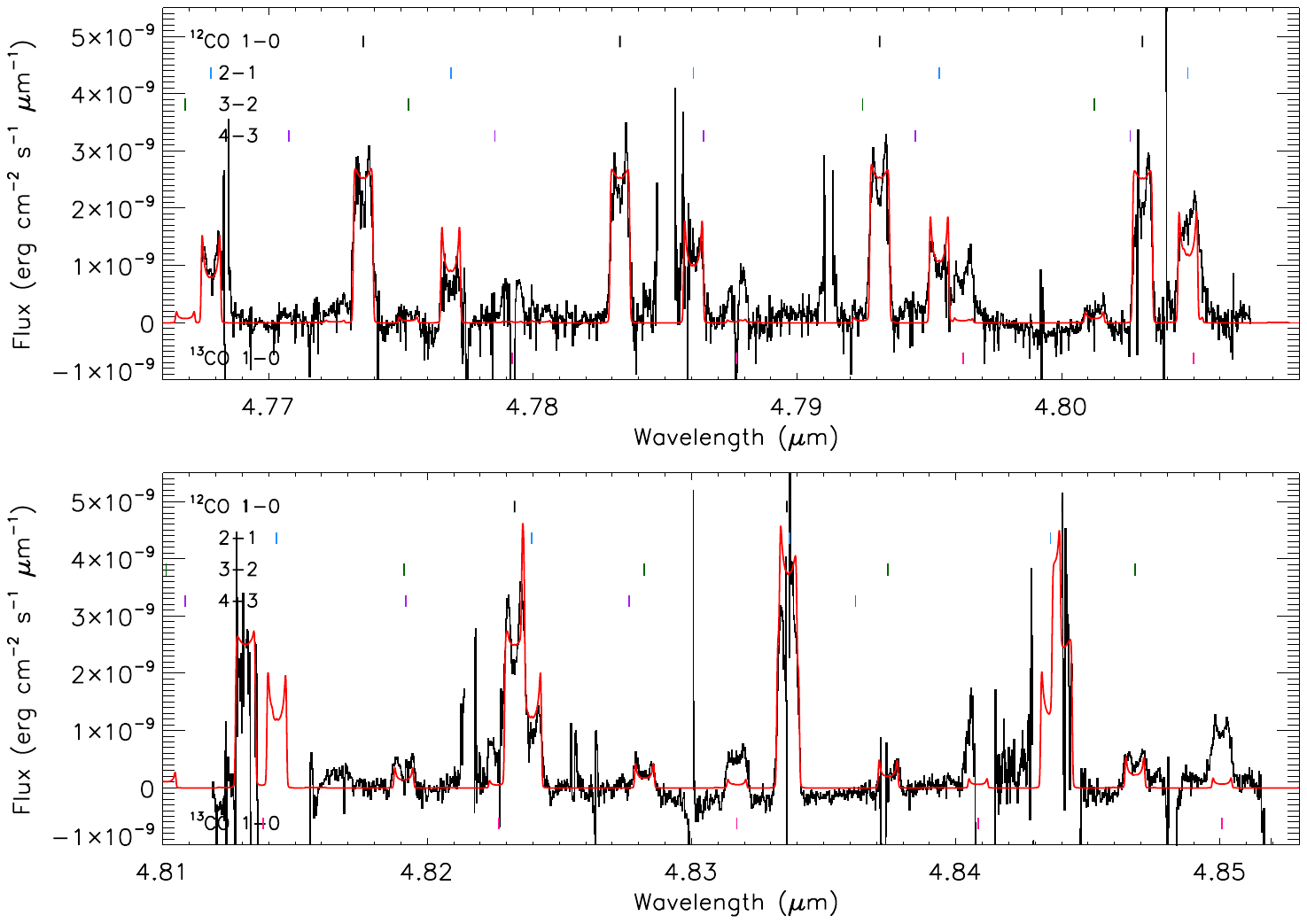}  
 \end{minipage}
 \begin{minipage}[]{8.3cm}{}
\includegraphics[width=6.9cm,angle=90]{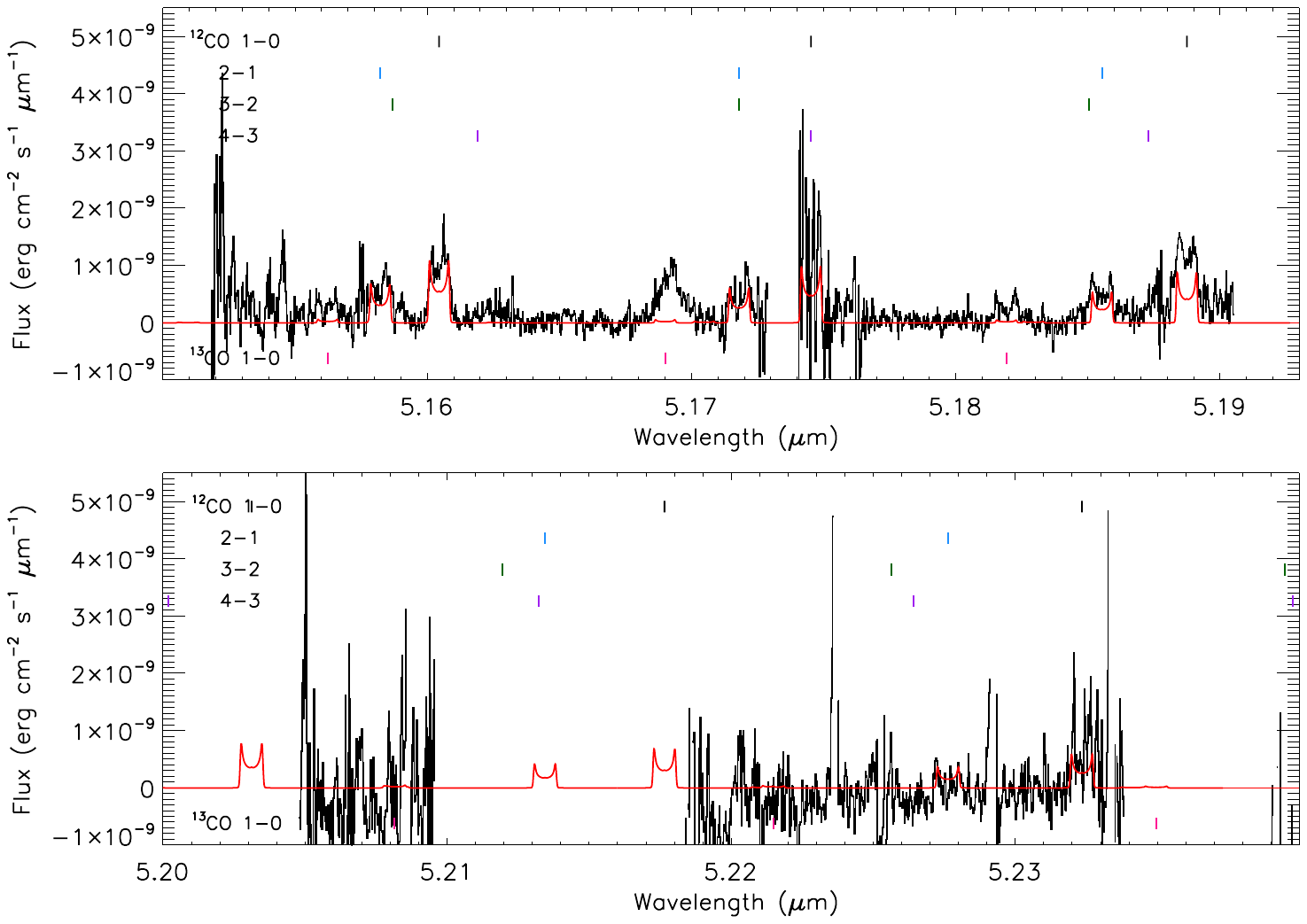}  
 \end{minipage}
\figcaption[]{The ring model for T = 1500 K
and a log density of 18.2 cm$^{-2}$ in red overlaid on the observed M band spectrum.  $^{13}$CO needs 
to be scaled by 6 to match the line profiles.
The line transitions are marked and the rovibrational bands they belong to are indicated at the 
top for $^{12}$CO and at the bottom shows $^{13}$CO 1 -- 0. Regions of
strong telluric absorption have been blanked out.
\label{fig-ringfit1}
}
\end{figure*}

Inspection of the K and M band spectra yields conflicting information regarding
the CO emitting gas. The relative strengths of the  $^{12}$CO lines across the M
band suggest that the gas is optically thick. In addition, the $^{13}$CO
emission lines are much stronger relative to those of $^{12}$CO than expected
given the normal isotopic ratio and optically thin gas. However, the clearly
double-peaked profiles suggest the emission cannot be completely optically
thick.

In order to determine the properties of the CO emitting gas, we constructed
models of the emission regions under the assumption that the gas was in
Keplerian orbit around a 10 $M_\odot$ central source. We considered two
configurations: a uniform (i.e., constant temperature and surface density) ring
and a thin disk whose temperature and surface density varied with radial
distance. We adopted an inclination of 55\degr\ derived by \citet{Vacca22}. The
models assume the gas is in LTE and are similar to those constructed by
\citet{Kraus00} and others. We generated a synthetic spectrum of the M band
emission for each value of temperature between 900 and 2000 K in steps of 100 K
and log CO surface density between 15.0 and 22.0 in steps of 0.2. We assumed a
source distance of 418 pc and the standard isotopic ratio for
$^{12}$CO/$^{13}$CO of 77 \citep{Wilson94}. For the ring models we adopted a
radius of 12 AU in order to match the mean width of the emission lines, and a
ring width of 1 AU. For the disk models we adopted a power law index of 0.4 for
the temperature and 1.5 for the surface density and radial steps of 0.5 AU
between an inner radius of 5 AU and an outer radius of 25 AU.

We then reddened each model by $A_V = 8.1$ mag, convolved it with a Gaussian
whose width is given by the instrumental resolution ($R \sim 88,000$ in the M
band), and sampled it at the observed wavelength sampling of the observations. A
$\chi^2$ comparison between the observed $^{12}$CO line spectra and the model
spectrum was then computed. For the ring model we computed the best fit scaling
of the model to the spectroscopic data, which corresponds to the best fit width
of the ring. The emission line strengths were also compared with the values
measured from the observed spectra on the standard rotational diagram
(Figure~\ref{fig-ladderfit}).

 Somewhat surprisingly, we found that no disk models with our assumed parameters
 were able to provide good matches to the observed spectra. A thin ring model
 with a temperature of $1500$ K and a surface density of $\log N_{CO} = 18.2$
 provided a reasonable, although not perfect, fit to the data. It was
 immediately clear, however, that the $^{13}$CO emission was far weaker in the
 model than in the actual data by a factor of 6 - 7. This was also reflected in an offset between the
 points for the $^{13}$CO emission line strengths from the model and that of the
 data on the rotational diagram, although the slope of the points for the
 $^{13}$CO emission line strengths from the model matched that of the data
 (which indicates that the derived temperature was approximately correct). This
 result, in addition to the differences seen in the line widths of the $^{12}$CO
 and $^{13}$CO emission profiles, suggests that the $^{12}$CO and $^{13}$CO
 emission arises from different parts of the disk. Therefore, we generated a
 pure $^{13}$CO emission model in which the $^{13}$CO gas temperature was the
 same as that for the $^{12}$CO gas, but the abundance was enhanced by a factor
 of 6 , which gave a good fit to the $^{13}$CO 1 -- 0 spectra.
 Figure~\ref{fig-ringfit1} shows the observed spectra overplotted with the best fit model without
 any additional scaling for $^{13}$CO.
 The corresponding line strengths are shown on the rotation diagram in Figure
 ~\ref{fig-ladderfit}.

Despite the excellent agreement between the model and the line strengths seen in
Figure~\ref{fig-ringfit1}, it is clear that the best-fit model is somewhat
unsatisfactory. A thin ring of CO emission seems physically unrealistic.
Furthermore, although the model provides a reasonable match to the overall line
strengths, it does not provide a good match to several prominent lines in the
spectrum. Many of the $^{12}$CO 2--1 lines in the model are much stronger than
in the data. In addition, the line profiles of many of the low lying $^{12}$CO
-- 0 transitions are not well reproduced by the model.

We have not done a detailed analysis of the $K$-band overtone spectrum, but the
 -- 0 bandhead emission is very weak  and the 3 -- 1 bandhead is almost
missing, indicating that the CO gas is quite cold. The best fit model from
the analysis of the rovibrational CO lines provides an adequate fit to the data. 

\subsection{Absorption lines}

The rovibrational fundamental CO lines in the low P and R transitions show
narrow absorption lines cutting out most of the emission from the accretion disk
in the center of the emission lines in the 
$M$ band spectra. Absorption is also
seen in the  lowest R and P transitions of $^{13}$CO in both data sets. This
suggest that the emission is being absorbed by a cold foreground cloud. This 
cold foreground cloud is likely to contribute to most of the extinction toward
MWC\,297, which is estimated to be $\sim$ 8.1 mag \citep{Vacca22}.

\begin{figure}[h]
\includegraphics[width=7cm,angle=90]{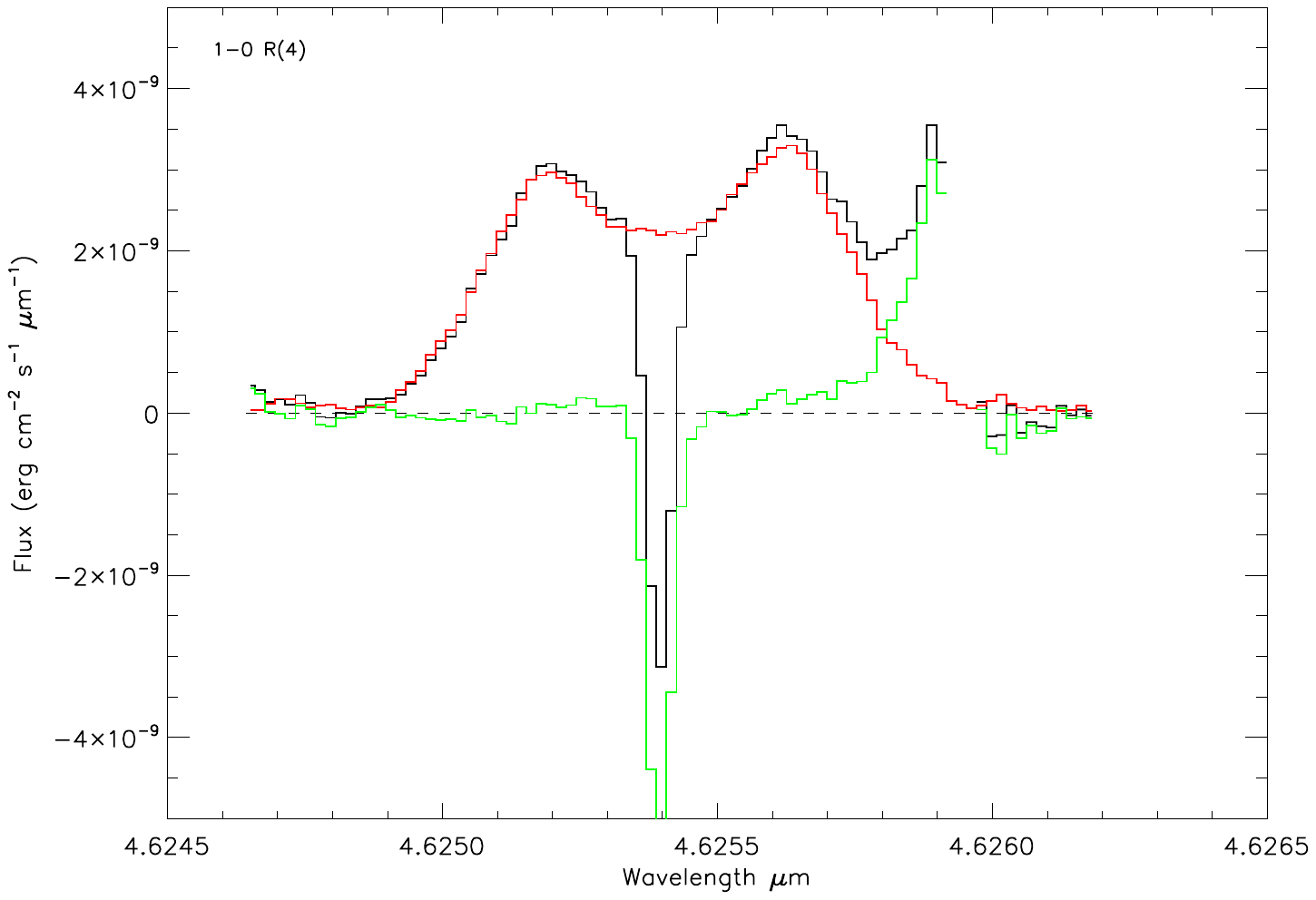}
\caption{An example illustrating the extraction of absorption line profiles.
The observed emission line, in this case  the $^{12}$CO $\upsilon$ = 1 -- 0 R(4) is plotted in black,
the scaled template profile is plotted in red and the resulting absorption profile is green.
The baseline around the absorption profile (green) is quite flat and centered at 0.
\label{fig-COsubtract}}
\end{figure}

\begin{figure}[h]
\includegraphics[width=7.5cm,angle=90]{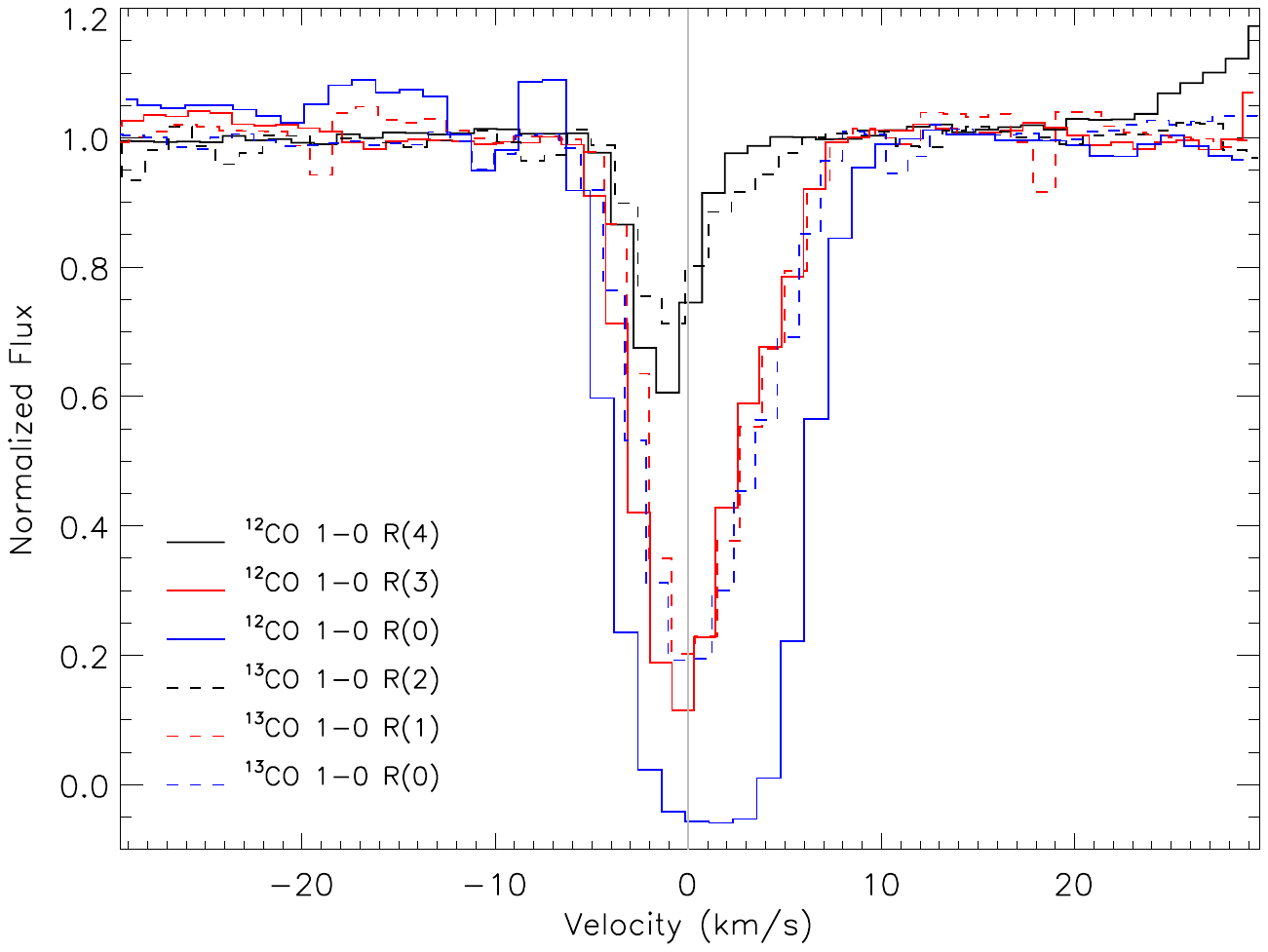} \caption{ $^{12}$CO and
 $^{13}$CO $\upsilon$ = 1 -- 0 absorption lines. The $^{12}$CO profiles are plotted
with solid lines and  $^{13}$CO with  dashed lines. The lines are identified in the bottom left corner
of the image. The velocity of the emission
lines, centered at 0 \kms, is marked with a gray vertical line. The optically thick $^{12}$CO
$\upsilon$ = 1 -- 0 R(0) line appears flat topped, much broader, and shifted in
velocity relative to $^{12}$CO 1 -- 0 R(4) and   $^{13}$CO 1 -- 0 R(2), which are
optically thin. This is because there is a second, colder red-shifted cloud
component, which is optically thick in the $^{12}$CO 1 -- 0 R(0) line. It is also seen
in  the  $^{13}$CO 1 -- 0 line as a strong red-shifted wing. The 0 \kms\ velocity corresponds to a V$_{lsr}$
= 6.7 \kms. \label{fig-COAbslines}}
\end{figure}

The lowest transitions of the CO  absorption lines in the 2020 data set are
strongly affected by telluric absorption,  which makes it difficult to estimate
the true extent of the absorption. Therefore, our analysis focussed on the April
 spectra, for which the relative velocity between the intrinsic absorption
and the telluric absorption is larger.   In order to extract the absorption
lines we generated a mean profile for the $\upsilon$ = 1 -- 0  $^{12}$CO
emission lines and then scaled that to match the observed profiles of the low J
transitions that have the foreground absorption (see
Figure~\ref{fig-COsubtract}). Examination of the CO absorption lines indicates
that the lowest transitions reach the zero intensity level and appear much
broader than the higher transitions, suggesting that they are optically thick.
The higher level J transitions agree well in velocity with the $^{13}$CO
absorption lines, while there is a clear shift in velocity at the lowest levels.
We illustrate this in Figure~\ref{fig-COAbslines}, which shows three CO
$\upsilon$ 1 -- 0 lines and three of the  $^{13}$CO lines in the source
spectrum, which has been shifted to 0 km/s. It can be seen from this figure that
the weaker absorption lines are blue-shifted relative to the disk emission. and
that the lines are asymmetric, with a red tail (Figure~\ref{fig-COAbslines}).
For the optically thin $^{12}$CO and  $^{13}$CO lines we find a velocity of -0.6
$\pm$ 0.5 \kms. With increasing optical depth  blue-shift of the lines
decreases and the red tail increases. At the lowest J levels the CO is very
optically thick, the profiles look flat topped and the center of the line is
shifted to the red. This suggests that there are two absorption components to
these lines, with the red shifted component contributing substantially to the
width of the lowest and most optically thick transitions. The radial velocity of
MWC\,297, 6.7 $\pm$ 2.2 \kms, which we determined for the center of the emission
lines is in good agreement with the velocity of the molecular cloud  in which it
was born.   \citet{Sandell23} determine a velocity $\sim$ 7.2 \kms\  of the
molecular cloud from $^{13}$CO(6--5) observations with APEX. This agrees well 
with $^{13}$CO(3-2) and C$^{18}O$(3--2) observations. These lines are unaffected
by the cold foreground cloud, while all the  emission in the low J $^{12}$CO
emission is completely absorbed \citep[see also][]{Manoj07}. Only emission from
CO(4--3) and higher rotational levels are unaffected by the cold foreground
cloud

To determine the temperature and column density of the absorbing gas we
performed a curve of growth analysis similar to that of  \citet{Goto03}. We
solved for column density, temperature, and the line width using an iterative
least-squares fitting routine. However, the low excitation CO lines are very
optically thick and also include the red-shifted component, which we cannot
separate at the iShell resolution. We therefore first analyzed the  weaker, and
presumably more optically thin, $^{13}$CO lines.  Using all  $^{13}$CO lines in
the analysis resulted in large densities (2 10$^{18}$ cm$^{-2}$) and very low
temperatures ($\sim$4 K). While these parameters provided a reasonable match the
observed equivalent widths (EWs), the model did not reproduce the line profiles
or the depths. The fit is skewed to high densities in order to reproduce the
observed EWs, which are clearly dominated by the second red absorption component
for the highest optical depth lines. As can be seen in
Figure~\ref{fig-COAbslines}, even the $^{13}$CO lines exhibit a red wing, which
is contributing to the EW. We therefore restricted the EW analysis to the two
lowest opacity lines, for which the red wing appears to have the smallest
contribution, and derived a column density of $\sim$ 6.5 10$^{17}$ cm$^{-2}$ and
a temperature of $\sim$  8.5 K. We then combined the two weakest CO lines with
the two weakest  $^{13}$CO lines in the fit and obtained a solution that matched
both the EWs and the line profiles. We kept the velocity width equal to 3.5 \kms
for this fit, which is in aqreement with our resolution (R $\sim$ 88,000). This
resulted in a CO column density, N(CO) = 6.7  $\pm$ 0.4  10$^{17}$ cm$^{-2}$,
and a temperature T = 8.25 $\pm$ 0.10 K.  No substantial residuals are seen when
we subtract this model  from the spectrum of the 4 lines. If we assume a CO
abundance of 10$^{-4}$, and that all the hydrogen is molecular and adopt
\begin{math} N(H_2)/A_V = 0.94~10^{21} \end{math} molecules cm$^{-2}$
\citep{Bohlin78}, then this absorption component indicates an extinction of 6.3
mag. This, however, is a lower limit to the total foreground extinction since
the red-shifted absorption component is completely opaque at the lowest CO
transitions and also detected in  $^{13}$CO. It is likely that it is responsible
for an additional extinction of 1 - 2 magnitudes. Therefore the cold foreground
CO emission readily explains the observed foreground extinction
of 8.1 mag.

\begin{figure}[t]
\includegraphics[width=7cm,angle=90]{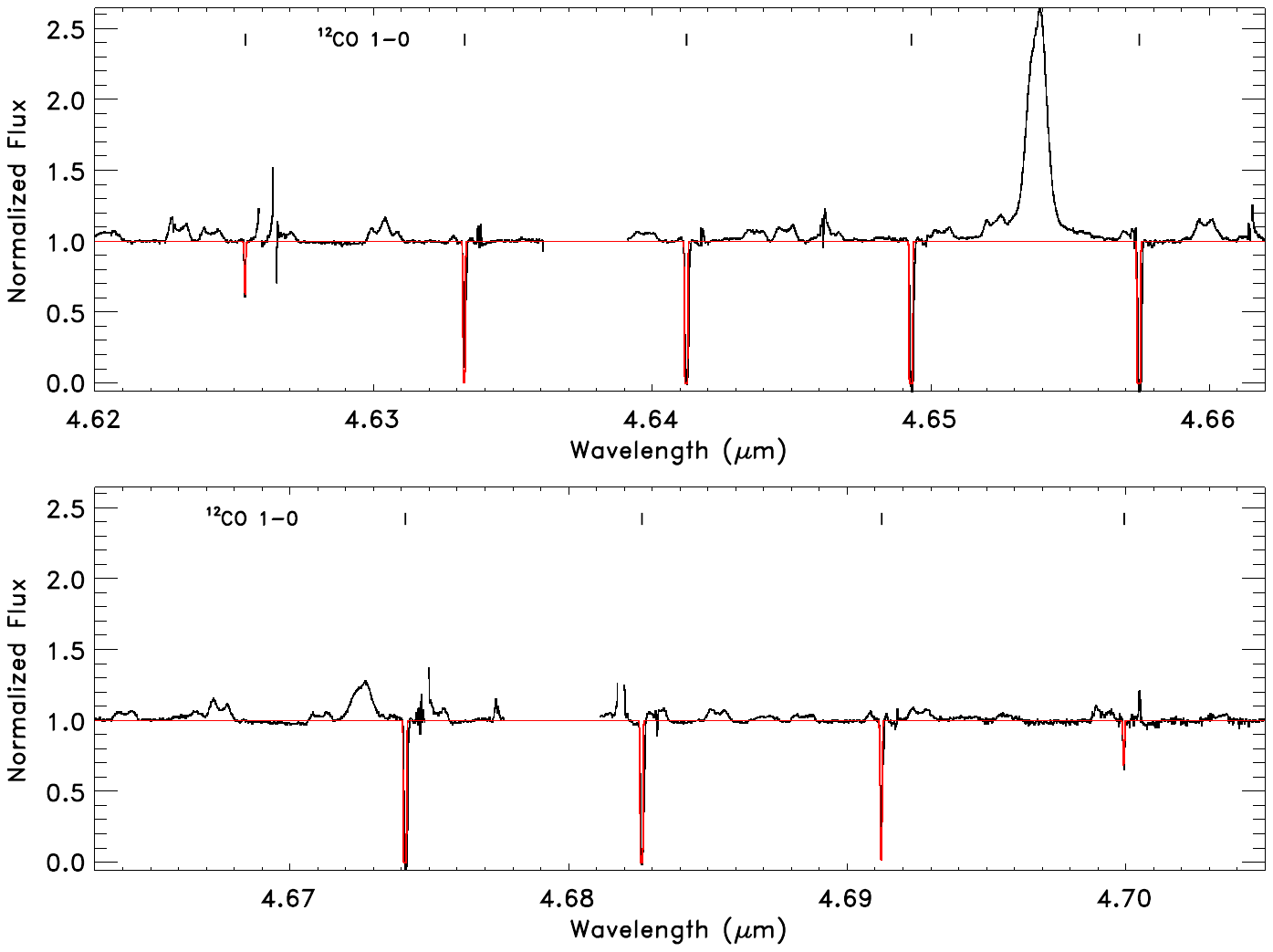}
\figcaption[]{
iSHELL  spectrum showing the $\upsilon$ = 1 -- 0 $^{12}$CO lines in absorption. The wavelength 
region is split into two for clarity. The transitions R(3), R(2), R(1) and R(0) at the top, P(1), P(2), 
P(3) and P(4) on the bottom, left to right. Regions of
strong telluric absorption have been blanked out. The fainter double peaked
emission lines are from higher rotational rotation transitions. The strong
hydrogen lines Pf $\beta$ and Hu $\epsilon$ are also seen. The model fit to the
absorption lines is overlaid in red.  
\label{fig-absorption}
 }  
\end{figure}

 \section{Discussion}
 
 Generally half of all solar mass stars have lost their disks after 3 Myr  with
an overall disk life time of 6 Myr \citep{Haisch01}. Some stars still have disks
after 10 Myr \citep[see e.g.][]{Espaillat08}. However, disk dispersion
timescales for high-mass stars are much shorter. By the time O and B stars
become optically visible, they have long since dispersed their disks. It is
therefore not surprising that hardly any disks have been detected around visible
early B stars. The two previously known examples are the early B  star
MWC\,349\,A, and LkH$\alpha$\,101. The latter is a B1 Ve star that illuminates
an \ion{H}{2} region. Infrared interferometry shows that it is surrounded by an
almost face-on disk with an inner hole and a dust radius of $\sim$ 30 AU
\citep{Tuthill02}. MWC\,297 is now a third example. It has long been suspected
to have an accretion disk, but the disk size and inclination have been
uncertain. MWC\,297 is  a much younger star than MWC\,349\,A or
LkH$\alpha$\,101, as it is still heavily accreting with  an age in the range of 
 -- 2.8 $\times$ 10$^5$ yr \citep{Fairlamb15,Vioque18}.

The rovibrational CO lines as well as the CO overtone bands show clear double
peaked profiles confirming that we see inclined circumstellar disk  around
MWC\,297, which is in Keplerian rotation.  MWC\,297 does not really fit in the
classification scheme by \citet{Meeus01} since it is much younger than typical
HAeBe stars, but if one follows the Meeus classification it would be in group I
\citep{Guzman21}. For this reason \citet{Acke04} added a group III for deeply
embedded and accretion dominated HAEBE stars. Neither does MWC\,297 readily fit
into the classification of CO spectra by \citet{Banzatti18,Banzatti22,
Bosman19}. It has double peaked profiles and the ratio of vibrational levels of
$\upsilon$2/$\upsilon$1 = 0.45 it in region 1, but since there is no cold dust
in the disk \citep{Vacca22}, we cannot definitely say whether it belongs to
region 1a or 1b. However, since the UV-fluorescence dominates the excitation of 
the $\upsilon$ = 2 -- 1 and  3 -- 2  bands and even appears to dominate the high
 excitation levels of CO 1 -- 0, it should probably go into group 1b. We also
note there are only five HAeBe stars that have overtone bandhead CO emission
\citep{Banzatti22}, yet the bandhead emission was  easily detected in MWC\,297.
Usually the overtone emission comes from hot, dense CO gas in the innermost part
of the disk \citep{Ilee13,Ilee14} with gas temperatures of $\sim$ 3000 K and
column densities of 10$^{21}$ cm$^{-2}$. The overtone emission in MWC\,297 is
much colder, $\sim$1500 K, an order of magnitude less dense, and extending up to
a radius of 10 AU.
 
 It was clear from the modeling of the CO 1 -- 0 lines that the line strengths
cannot be fit with a single temperature model. The is very little change in line
strength between low and high excitation lines and they all have the
characteristic double peaked profile from a disk in Keplerian rotation. An
isothermal fit to all the rotational lines results in very high optical depths
for the lowest transition and makes the line profiles flat-topped, which is not
what is observed. It is clear that the lines can only be moderately optically
thick, since they all show a clear U-shaped profile. Neither can a single
temperature model explain the strong $\upsilon$ = 2 - 1 and 3 -- 2 transitions,
which for the 2 -- 1 lines are almost half the line strengths of the 1 -- 0
lines. This was already noticed by \citet{Brittain03}, who argued that the
strong $\upsilon$ = 2 - 1 and 3 -- 2 transitions in isolated HAeBe star
HD\,141569 must be excited by UV fluorescence, see also \citet{Brittain07}.
Recent modeling of  HD\,141569, which is a well studied HAeBe star with
vibrational levels detected  up to $\upsilon$ = 7 -- 6 confirm that fluorescence
is needed to explain the observed rovibrational lines \citep{Jensen21} We expect
MWC\,297 to  have strong UV radiation, both from the fact that it is an early B
star and because it is heavily accreting. However, due to the high foreground
extinction, 8 .1 mag, the star is not visible in the UV. Therefore we have not
attempted to include  fluorescence in ir model, although it it is clear that UV
fluorescence is needed to explain the emission in the higher vibrational bands
and even necessary for the high excitation 1 -- 0 lines. The same may be true
for other HAeBe disks as well.

Because MWC\,297 is very bright in the infrared, it has been the target of many
interferometric studies. These studies show a well resolved inner disk in the
continuum with a size of  2 -- 4 AU in continuum 
\citep{Malbet07,Kraus08,Acke08,Weigelt11,Hone17, Kluska20} with the line
emission (Br$\gamma$) being 2 -- 3 times more extended
\citep{Malbet07,Kraus08,Weigelt11,Hone17}. \citet{Acke08} also observed MWC\,297
in the mid-inrared. Modeling of mid-infrared data required a two component model
and they found that a smaller inner disk with a size of  4 AU and a more
extended disk with a size of 17 AU (40 mas), This is rather similar to the size
of  disk we see in CO Neither do these studies find any inner gap
\citep{Acke08,Kluska20}. What is also noteworthy is that the size of the
continuum region is well within the dust destruction radius, which for MWC\,297
is $\sim$ 5 AU\footnote{\citet{Kluska20} quotes a dust destruction radius of 8
AU, but that is because they use a stellar luminosity  4 $\times$ 10$^4$ \Lsun,
which is far too high for a B1.5 V star, see \citet{Vacca22}.}. As pointed out
by \citet{Vacca22} the Br$\gamma$ emission is extremely optically thick and it
is therefore appears that the continuum comes from an optically thick gaseous
region. From the observed line width of the CO lines at 10\% level,  $\sim$ 65.2
\kms, we find the inner CO radius $\sim$ 5.6 AU, i.e.
roughly at the same radius as the dust destruction radius,

A normal isotope ratio of [$^{12}$C/$^{13}$C] of 77 greatly underestimates the
strength of the observed $^{13}$CO line intensities. We find that a relative
$^{12}$CO to $^{13}$CO  abundance of $\sim$ 6 provides a good fit to the data.
This does not mean that the disk has anomalous isotope ratios, but rather that
$^{13}$CO is not in LTE and coexisting with $^{12}$CO. As we seen, the FWHM of
$^{13}$CO 1 -- 0 differs by more than 5\% from that of 1 -- 0, see
Section~\ref{sect-Analysis}. These kind of $^{12}$CO/$^{13}$CO ratios are not
uncommon. \citet{Plas15} found that the $^{12}$CO/$^{13}$CO ratio varied between
 and 7 for their sample of 12 HAeBe stars and that the rotational temperatures
of $^{13}$CO were on the average lower than those for $^{12}$CO. 
\citet{Banzatti22}, who has the largest sample of rovibrational CO lines in
HAeBe stars, do not quote a $^{12}$CO/$^{13}$CO ratio, but looking at the line
ratios for their sample of 17 HAeBe stars with double peaked line profiles shows
that the ratio ranges from 3 to 25, with a median of 14. This does not directly
translate  to an abundance ratio  [$^{12}$CO/$^{13}$CO],  because for most disks
$^{12}$CO 1 -- 0 is more optically thick than $^{13}$CO 1 --0 , nor does it account for
difference in excitation between $^{12}$CO and $^{13}$CO.

We detect narrow absorption lines, FWHM $\sim$ 4 - 5 \kms\ from a cold unrelated
foreground cloud in both $^{12}$CO and  $^{13}$CO. Radio observations show that
this foreground cloud is very extended and therefore more local, probably
somewhere  around 100 to 200 pc from us. At the lowest J transitions the
absorption lines  in  $^{12}$CO are completely saturated. At higher J levels the
lines become more blue-shifted and agree in velocity with the $^{13}$CO
absorption lines. The lowest transitions of $^{13}$CO is also optically thick
and show a red-shifted line wing.wing.  Detailed analysis shows that the
red-shifted component comes from a  colder, but less dense cloud component. For
the blue-shifted cloud component, which is at a V$_{lsr}$  $\sim$ 6.4 \kms\ we
derive a temperature of  8.3 $\pm$ 0.1 K and a  CO column density, N(CO) = 6.7
$\times$  10$^{17}$ cm$^{-2}$. If we assume that all the CO gas is molecular
with a typical abundance ratio of 10$^{-4}$ it corresponds to an extinction of
.3 magnitude assuming the standard relation between molecular hydrogen and
extinction. Since the redshifted cloud component also contributes to the
extinction, it is clear that foreground cloud plus diffuse gas along the line of
sight can fully explain the observed high extinction, 8.1 mag, toward MWC\,297.
\citet{Banzatti22}  noticed that  the absorption line  FWHM decreased as a
function of J level from 10 \kms\ in the J = 1 lines down to a minimum of 3.3
\kms\ in the $\upsilon$ = 1 -- 0 J = 4 lines and that the line shape also
changed, becoming less Gaussian and showing effects of saturation at the lowest
J levels. They also saw a similar behavior in $^{13}$CO. This agrees well with
what we have found, and we now now it is due to two cold clouds in the
foreground of MWC\,297 with different temperatures and densities, but both of
which are completely optically thick in $^{12}$CO at the lowest J levels.

\section{Summary and Conclusions}

We have shown that  overtone bandhead and rovibrational CO lines trace a hot gas
disk in Keplerian rotation around the 10 \Msun\ star MWC\,297. Our modeling
shows that the emission cannot be explained by thermal excitation alone. It
therefore appears that the high excitation rotational lines are largely excited
by fluorescence, something which has been seen in other HAeBe disks as well. 
Analysis of the spectra show that $^{12}$CO $\upsilon$ = 1--0 emission is
optically thick for the low excitation lines. Even the $^{13}$CO 1 -- 0 and 
$^{12}$CO 2 -- 1 have somewhat optically thick lines at low J levels. We find
that a narrow ring with a radius of 12 AU and an inclination of 55\degr\ 
provides an adequate fit to data. For this model we derive a rotational
temperature of 1500 K  and a CO column density of 1.6 $\times$ 10$^{18}$ 
cm$^{-2}$.  This model underestimates the line strength of high J lines,
indicating that these lines are excited by fluorescence. The CO overtone
emission, which is only seen in a few HAeBe disks, has a similar temperature.
Our best-fit model is still somewhat unsatisfactory. A thin ring of CO emission
seems physically unrealistic. Furthermore, although the model provides a
reasonable match to the overall line strengths, it does not provide a good match
to several prominent lines in the spectrum. Many of the $^{12}$CO 2--1 lines in
the model are much stronger than in the data.  We also find that a normal
isotope ratio of [$^{12}$C/$^{13}$C] of 77 greatly underestimates the $^{13}$CO
line intensities, while a $^{12}$CO to $^{13}$CO  abundance of $\sim$  13
provides a good fit to the data. Other HAeBe disks typically show similarly
enhanced  $^{13}$CO line intensities, most likely because excitation conditions
in the disks differ between $^{12}$CO and $^{13}$CO. 

The $^{12}$CO 1 -- 0 spectrum shows a strong absorption feature in the center of
the emission profile, which is completely saturated at J = 1 and 2, and which is
also seen in  $^{13}$CO 1 -- 0. This absorption is due to cold foreground
emission, which is unrelated to MWC\,297. Radio observations show that this
foreground cloud is very extended and therefore more local, probably somewhere 
around 100 to 200 pc from us. Detailed analysis show that this foreground
absorption is caused by two cold foreground clouds, partly overlapping in
velocity. The more opaque one is slightly blue-shifted relative to the center of
the emission lines has a temperature of 8.3 K and a column density of 6.7
$\times$  10$^{17}$ cm$^{-2}$, which correspond to a visual extinction of  6.3
mag. With our velocity resolution we could not separate the second red-shifted
component well enough to be able to good estimate the temperature and density,
but it is apparent that it is colder and has a lower column density. It is clear
these foreground clouds are responsible for the high extinction toward MWC\,297.

 We have shown that the young, heavily accreting B1.5 V star MWC\,297 is still
 surrounded by a molecular accretion disk in Keplerian rotation. It is the only
 early B star, which has been detected in ro-vibrational CO lines and one of the
 few HAeBe stars detected in overtone CO emission.  The circumstellar disk has
 not been detected at  mm-wavelengths, because  a cold foreground cloud absorbs
 all the low J $^{12}$ CO lines. These lines are typically used for detecting
 gas in disk with facilities like SMA and ALMA. The MWC\,297 disk does not have
 any cold dust. It that sense it resembles MWC\,349A, which has been detected in
 CO overtone emission but has not been detected in any molecular line at
 mm-wavleengths, suggesting that it has no cold molecular gas. For the same
 reason we find it very unlikely that MWC\,297 would have any cold molecular
 gas. Based on the rovibrational lines it also looks like the disk is rather
 compact. It would be very interesting to get a better estimate of the size of
 the disk using spectroastrometric imaging using the rovibrational CO lines.\\

We appreciate Prof.  Andrea Banzatti sharing some of his iSHELL data on MWC\,297
with us. We also thank Dr. Adwin Boogert  for helping us with the telluric
correction of the September 2020 iSHELL spectrum. We are especially grateful to
the anonymous referee, whose constructive criticism encouraged us to completely
rethink our modeling approach to the CO lines and to get additional data.


\end{document}